\documentclass[journal, hidelinks]{IEEEtran}
\usepackage[switch]{lineno}

\usepackage{cite}
\usepackage{amsmath,amssymb,amsfonts,mathtools}
\usepackage{algorithmic}
\usepackage{graphicx}
\usepackage{textcomp}
\usepackage{xcolor}
\usepackage{orcidlink}
\def\BibTeX{{\rm B\kern-.05em{\sc i\kern-.025em b}\kern-.08em
		T\kern-.1667em\lower.7ex\hbox{E}\kern-.125emX}}
\usepackage[final]{microtype}
\usepackage{gensymb}

\usepackage[toc, acronym]{glossaries}
\glsdisablehyper

\usepackage{tabto}
\usepackage{comment}
\usepackage{siunitx}
\usepackage{graphicx}
\graphicspath{{figures/}}

\usepackage{pgfplots}
\pgfplotsset{compat=newest}
\usepgfplotslibrary{groupplots}
\usepgfplotslibrary{polar}
\usepgfplotslibrary{smithchart}
\usepgfplotslibrary{statistics}
\usepgfplotslibrary{dateplot}
\usepgfplotslibrary{ternary}
\usetikzlibrary{arrows.meta}
\usetikzlibrary{backgrounds}
\usepgfplotslibrary{patchplots}
\usepgfplotslibrary{fillbetween}

\let\originalleft\left
\let\originalright\right
\renewcommand{\left}{\mathopen{}\mathclose\bgroup\originalleft}
\renewcommand{\right}{\aftergroup\egroup\originalright}

\usepackage[nameinlink,capitalize]{cleveref}
\crefname{figure}{Fig.}{Figs.}
\crefname{equation}{}{}
\crefname{section}{Sec.}{Secs.}

\newcommand{\herm}[1]{\ensuremath{#1^{\mathrm{H}}}}
\newcommand{\col}{\ensuremath{\mathrm{col}}}
\newcommand{\euler}{\ensuremath{\mathrm{e}}}
\newcommand{\im}{\ensuremath{\mathrm{j}}}
\newcommand{\norm}[2]{\ensuremath{\|#1\|_{#2}}}
\newcommand{\cnumbers}{\ensuremath{\mathbb{C}}}
\newcommand{\complexnormal}[2]{\ensuremath{\mathcal{CN}(#1, #2)}}
\newcommand{\expec}[1]{\ensuremath{\mathrm{E}_{#1}}}
\newcommand{\trans}[1]{\ensuremath{#1^{\mathrm{T}}}}
\newcommand{\rk}[1]{\ensuremath{\mathrm{rank}(#1)}}
\newcommand{\id}[1]{\ensuremath{\mathbf{I}_{#1}}}
\newcommand{\zeros}[1]{\ensuremath{\mathbf{0}_{#1}}}
\newcommand{\tr}[1]{\ensuremath{\mathrm{tr}\left(#1\right)}}
\newcommand{\perr}{\ensuremath{P_{\epsilon}}}
\newcommand{\bsf}[1]{\ensuremath{\boldsymbol{\mathsf{#1}}}}
\newcommand{\cov}[1]{\ensuremath{\mathbf{R}_{#1}}}
\newcommand{\vect}{\ensuremath{\mathrm{vec}}}
\newcommand{\dist}[1]{\ensuremath{d_{\mathrm{#1}}}}
\DeclareMathOperator*{\argmin}{arg\,min}

\newcommand{\ie}{\textit{i.e.} }
\newcommand{\eg}{\textit{e.g.} }

\IEEEaftertitletext{\vspace{-1\baselineskip}}

\usepackage{xcolor}
\usepackage{fancyhdr}
\newcommand{\doi}{10.1109/LWC.2025.3572761}
\newcommand{\publication}{IEEE Wireless Communications Letters}
\newcommand{\changefont}{
	\color{blue}\fontsize{9}{9}\selectfont
}

\let\oldmaketitle\maketitle
\renewcommand{\maketitle}{%
	\oldmaketitle
	\thispagestyle{fancy} % Title page has fancy
}

\begin{document}
	\glsdisablehyper
\newacronym{rf}{RF}{radio frequency}
\newacronym{mimo}{MIMO}{multiple-input multiple-output}
\newacronym{bs}{BS}{base station}
\newacronym{ue}{UE}{user equipment}
\newacronym{upa}{UPA}{uniform planar array}
\newacronym{csi}{CSI}{channel state information}
\newacronym{nlos}{NLoS}{non-line-of-sight}
\newacronym{ser}{SER}{symbol error probability}
\newacronym{pam}{PAM}{pulse-amplitude modulation}
\newacronym{snr}{SNR}{signal-to-noise ratio}
\newacronym{sinr}{SINR}{signal-to-interference-plus-noise ratio}
\newacronym{ml}{ML}{maximum likelihood}
\newacronym{3g}{3G}{third generation}
\newacronym{4g}{4G}{fourth generation}
\newacronym{5g}{5G}{fifth generation}
\newacronym{6g}{6G}{sixth generation}
\newacronym{thz}{THz}{terahertz}
\newacronym{dof}{DoF}{degrees of freedom}

	\bibliographystyle{IEEEtran-normspace}
	\bstctlcite{IEEEexample:BSTcontrol}

	\title{Harnessing Wavefront Curvature and Spatial Correlation in Noncoherent MIMO Communications}
	\author{Aniol Martí\,\orcidlink{0000-0002-5600-8541},~\IEEEmembership{Graduate~Student~Member,~IEEE},
		Luca Sanguinetti\,\orcidlink{0000-0002-2577-4091},~\IEEEmembership{Fellow,~IEEE},
		Meritxell~Lamarca\,\orcidlink{0000-0002-8067-6435},~\IEEEmembership{Member,~IEEE},
		and Jaume Riba\,\orcidlink{0000-0002-5515-8169},~\IEEEmembership{Senior~Member,~IEEE}%%
		\thanks{This work was supported by project MAYTE (PID2022-136512OB-C21) by MICIU/AEI/10.13039/501100011033 and ERDF/EU, grant 2021 SGR 01033 by Departament de Recerca i Universitats de la Generalitat de Catalunya and grant 2023 FI ``Joan Or\'o'' 00050 by Departament de Recerca i Universitats de la Generalitat de Catalunya and the\nobreak\ ESF+. L. Sanguinetti was supported in part by the Italian Ministry of Education and Research (MUR) in the framework of the FoReLab Project (Department of Excellence) and in part by the European Union under the Italian National Recovery and Resilience Plan (NRRP) of NextGenerationEU, partnership on ``Telecommunications of the Future'' (PE00000001 -- Program ``RESTART'', Structural Project 6GWINET, Cascade Call SPARKS).}%
		\thanks{A. Martí, M. Lamarca and J. Riba are with the Departament de Teoria del Senyal i Comunicacions, Universitat Politècnica de Catalunya (UPC), 08034 Barcelona (e-mail: aniol.marti@upc.edu, meritxell.lamarca@upc.edu, jaume.riba@upc.edu).}
		\thanks{L. Sanguinetti is with the Dipartimento di Ingegneria dell'Informazione, University of Pisa, 56122 Pisa, Italy (e-mail: luca.sanguinetti@unipi.it).}
	}

	\maketitle

	\begin{abstract}
        Noncoherent communication systems have regained interest due to the growing demand for high-mobility and low-latency applications.
        Most existing studies using large antenna arrays rely on the far-field approximation, which assumes locally plane wavefronts.
        This assumption becomes inaccurate at higher frequencies and shorter ranges, where wavefront curvature plays a significant role and antenna arrays may operate in the radiative near field.
        In this letter, we adopt a model for the channel spatial correlation matrix that remains valid in both near and far field scenarios.
        Using this model, we demonstrate that energy-based noncoherent systems can leverage the benefits of wavefront spherical curvature, even beyond the Fraunhofer distance, revealing that the classical far-field approximation may significantly underestimate system performance.
        Moreover, we show that large antenna arrays enable the multiplexing of various users even with a noncoherent processing, as well as permitting near-optimal detection with low computational complexity.
	\end{abstract}


	\section{Introduction}
        \IEEEPARstart{L}{arge} arrays have become a fundamental piece of current and next generation communication systems.
        Compared to traditional \acrfull{mimo}, large arrays exhibit several benefits such as the reduction of small-scale fading and noise, a phenomenon known in literature as channel hardening~\cite[Sec.~1.3]{marzetta_fundamentals_2016}.
        In order to fully exploit channel hardening, the receiver typically needs to obtain instantaneous \acrfull{csi} through a training sequence and use it for coherent detection of the transmitted data.
        However, in high-mobility environments and systems with a large number of users, the training overhead can result in significant performance degradation~\cite{chafii_twelve_2023,jing_design_2016}.
        A promising alternative is a paradigm shift to noncoherent communications, in which neither transmitter nor receiver have instantaneous \acrshort{csi} but only statistical.

        Large arrays are typically deployed in high frequency systems, as it allows to place more antennas in less space.
        However, this combination breaks a common assumption of communications: the wavefront at the receiver is planar~\cite{bacci_spherical_2023}.
		The electromagnetic field between a transmitting and a receiving antenna has been usually divided in three regions: the reactive near field, the radiative near field\footnote{Throughout this paper we only consider the radiative near field and refer to it just as \textit{near field}.} and the far field~\cite[Sec.~2.2.4]{balanis_antenna_2016}.
		The reactive and radiative near field are separated by the Fresnel distance, $\dist{r} = 0.62\sqrt{D^3/\lambda}$, whereas the near and far field are divided by the Fraunhofer distance, $\dist{F} = 2D^2/\lambda$, with $D$ the aperture and $\lambda$ the wavelength.

		The origin of the far-field approximation is that at the Fraunhofer distance the total phase error at an antenna due to the wavefront curvature is $\pi/8$, and from an antenna theory perspective angles smaller than this can be neglected~\cite[Sec.~4.4.1]{balanis_antenna_2016}.
		But, what if the spherical wavefront is assumed beyond the Fraunhofer distance?
        Indeed, \cite{bacci_spherical_2023} showed that the classical far-field approximation underestimates spectral efficiency in coherent multiuser \acrshort{mimo}, indicating that wavefront curvature can be exploited beyond the Fraunhofer distance.
        However, does this also hold for noncoherent systems?
        The main goal of this paper is demonstrating that the Fraunhofer distance neither is the appropriate boundary for the near and far fields in noncoherent communications.
        Furthermore, we show that multiuser noncoherent systems can leverage large arrays to multiplex different users and achieve near-optimal detection with lower complexity using a single-user detector.
        This suggests that simple constellations as those explored in~\cite{marti_constellation_2024} can be used in multiuser scenarios with minimal penalty if the number of antennas is sufficiently large~\cite{cuevas_advanced_2024}.

	\section{System and Signal Model}
		\label{sec:near-field_model}
\begin{figure}
				\centering
				\def\svgwidth{0.55\columnwidth}
				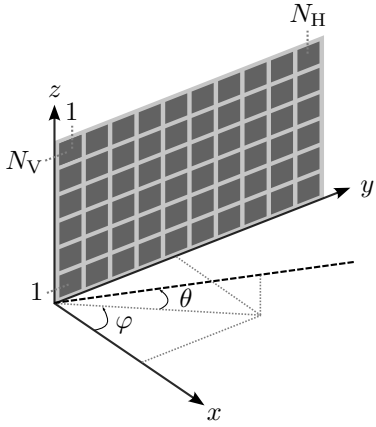
				\caption{Geometry of a \acrshort{bs} equipped with a $N_{\mathrm{H}} \times N_{\mathrm{V}}$ \acrshort{upa} in the $yz$ plane.}
				\label{fig:upa}
		\end{figure}

        We consider a \acrshort{mimo} system with $K$ active single-antenna \acrfullpl{ue} operating in \acrfull{nlos} propagation conditions. The \acrfull{bs} is equipped with a \acrfull{upa} located in the $yz$ plane, and consisting of $N$ independent \acrfull{rf} chains, arranged into $N_{\mathrm{V}}$ rows, each with $N_{\mathrm{H}}$ antennas: $N=N_{\mathrm{H}}N_{\mathrm{V}}$, as depicted in~\cref{fig:upa}.
        The inter-element spacing across horizontal and vertical directions is $\Delta$.
        The location in Cartesian coordinates of the $n$-th antenna, with $1 \le n \le N$, with respect to the origin is $\mathbf{u}_n=\trans{(0, i_n\Delta, j_n\Delta)}$, where $i_n = \mathrm{mod}(n-1, N_{\mathrm{H}})$ and $j_n = \lfloor\frac{n-1}{N_{\mathrm{H}}}\rfloor$ are the horizontal and vertical indices of element $n$.
        We call $\bsf{h}_k\in \mathbb{C}^N$ the channel response between the $k$-th \acrshort{ue} and the \acrshort{bs}~\cite{demir_spatial_2024}.

        \subsection{Channel Model}
            Under \acrshort{nlos} conditions, the \acrshort{bs} receives a superposition of $L$ multipath components~\cite[Sec.~1.3]{marzetta_fundamentals_2016}.
            With scattering localized around each \acrshort{ue} and the elevated \acrshort{bs} experiencing no near-field scattering, each component produces a wave arriving at the array from a specific direction~\cite[Sec.~2.6]{bjornson_massive_2017}.
            Under rich scattering, the channel vector follows a complex Gaussian distribution, $\bsf{h}_k \sim \complexnormal{\zeros{N}}{\cov{\bsf{h}_k}}$, with covariance matrix
    		\begin{equation}
    			\cov{\bsf{h}_k} = \expec{}[\bsf{h}_k\herm{\bsf{h}_k}] = \sum_{i=1}^{L} \beta_{ki}\mathbf{a}_{ki}\herm{\mathbf{a}_{ki}}.
    			\label{eq:correlation}
    		\end{equation}
            Here, $\beta_{ki}$ represents the path gain associated with the $i$-th scatterer and
    		\begin{equation}
    			\mathbf{a}_{ki}=\trans{\left(\euler^{-\im\frac{2\pi}{\lambda}\norm{\mathbf{s}_{ki}-\mathbf{u}_0}{}}, \dots, \euler^{-\im\frac{2\pi}{\lambda}\norm{\mathbf{s}_{ki}-\mathbf{u}_{N-1}}{}}\right)}
    		\end{equation}
            is the array response vector for the wave arriving from the scatterer located at $\mathbf{s}_{ki} = \trans{(r_{ki}, \theta_{ki}, \varphi_{ki})}$, with $r_{ki}$ being the radial distance, $\theta_{ki}$ the elevation and $\varphi_{ki}$ the azimuth, as defined in~\cref{fig:upa}.

            Unlike the classical model, which assumes locally planar wavefronts and only applies in the far field, the above model is valid in the so-called radiative near field of the array~\cite{dong_near-field_2022}.
            If the $k$-th \acrshort{ue} is in the far field of the array, then $\bsf{h}_k$ can be approximated as $\bsf{h}_k^{\mathrm{FF}} \sim \complexnormal{\zeros{N}}{\cov{\bsf{h}_k}^{\mathrm{FF}}}$, with $\cov{\bsf{h}_k}^{\mathrm{FF}}$ given by:
        	\begin{equation}
                \cov{\bsf{h}_k}^{\mathrm{FF}} = \sum_{i=1}^{L} \beta_{ki}\mathbf{a}_{ki}^{\mathrm{FF}}\herm{({\mathbf{a}_{ki}^{\mathrm{FF}}})},
		          \label{eq:correlation_ff}
		      \end{equation}
            where the $n$-th entry of $\mathbf{a}_{ki}^{\mathrm{FF}}$ can be computed as~\cite{demir_spatial_2024}:
			\begin{equation}
				[{\mathbf{a}_{ki}^{\mathrm{FF}}}]_{n} = \euler^{\,\im\frac{2\pi}{\lambda}\Delta(i_n\cos\theta_{ki}\sin\varphi_{ki}+j_n\sin\theta_{ki})}.
			\end{equation}

    	\subsection{Signal Model}
            Consider the uplink of a \textit{one-shot} communication system where the transmitted symbols are drawn from an equiprobable $M$-ary constellation $\mathcal{X} = \{x_1, \dots, x_M\}$ with $M\geq2$ and an average power constraint $\expec{}[|\mathsf{x}|^2]=1$.
            Since the channel is constant for a single use and information is decoded symbol-by-symbol, modulation schemes such as differential encoding have to be discarded~\cite[Ch.~12]{hanzo_mimoofdm_2010}.
    		The complex baseband signal $\bsf{y} \in \cnumbers^N$ received at the \acrshort{bs} is:
            \begin{equation}
                \label{eq:multiuser_model}
                \bsf{y} = \sum_{k=1}^{K}\bsf{h}_k\sqrt{p_k}\mathsf{x}_k + \bsf{z},
            \end{equation}
            where $\bsf{z}\sim\complexnormal{\zeros{N}}{\cov{\bsf{z}}}$ is additive Gaussian noise, $\mathsf{x}_k$ is the \acrshort{pam} symbol transmitted by the $k$-th \acrshort{ue} and $p_k$ is a power control factor such that at the \acrshort{bs} all signals have the same \acrfull{sinr},
            \begin{equation}
                \mathrm{SINR} = \frac{p_k\tr{\cov{\bsf{h}_k}}}{\tr{\cov{\bsf{z}}}+\sum_{j\neq k}p_j\tr{\cov{\bsf{h}_j}}}.
            \end{equation}
            As with any noncoherent communication system, the \acrshort{bs} is unaware of the realizations of $\{\bsf{h}_k\}_{k=1}^K$.
            On the other hand, it may rely on their statistical properties.

            Letting $\mathcal{X}^K$ be the $K$-ary Cartesian product of $\mathcal{X}$, the \acrfull{ser} is:
            \begin{equation}
                \label{eq:pe_multiuser}
                \perr = \frac{1}{M^K}\!\sum_{\mathbf{x}\in\mathcal{X}^K}\!\Pr[\hat{\mathbf{x}}(\mathbf{y}) \neq \mathbf{x} | \bsf{x} = \mathbf{x}],
            \end{equation}
            where $\hat{\mathbf{x}}(\mathbf{y})$ is the output of a multiuser symbol detector applied to the received signal $\mathbf{y}$.
            In particular, the optimal $\hat{\mathbf{x}}(\mathbf{y})$ (\ie the one that minimizes $\perr$) is the \acrfull{ml} detector~\cite[Sec.~4.1-1]{proakis_digital_2008}.
            This yields
            \begin{equation}
                \label{eq:ml_mu}
                \hat{\mathbf{x}}(\mathbf{y}) = \hat{\mathbf{x}}_{\mathrm{ML}} = \argmin_{\mathbf{x}\in\mathcal{X}^K} \herm{\mathbf{y}}\cov{\bsf{y}|\mathbf{x}}^{-1}\mathbf{y} + \log|\cov{\bsf{y}|\mathbf{x}}|,
            \end{equation}
            where
            \begin{equation}
                \label{eq:ml_correlation}
                \cov{\bsf{y}|\mathbf{x}} = (\trans{\mathbf{x}}\otimes\id{N})\cov{\vect(\bsf{H}\mathbf{P})}(\mathbf{x}^*\otimes\id{N}) + \cov{\bsf{z}},
            \end{equation}
            with $\mathbf{x} = \trans{(x_1, \dots, x_K)}$, $\bsf{H} \in\cnumbers^{N \times K} = (\bsf{h}_1 , \dots , \bsf{h}_K)$ and $\mathbf{P} = {{\rm diag}(p_1, \ldots,p_K)}$ the diagonal power control matrix.
            The column-wise vectorization is denoted by $\vect(\cdot)$ and $\otimes$ is the Kronecker product.

            Focusing on the previous expressions, \cref{eq:ml_mu,eq:ml_correlation} evidence that one-shot noncoherent schemes like those considered in this work are not able to retrieve symbol phase information, since the optimal detector can only perceive its energy~\cite{VilaInsa2024}.
            Furthermore, the capacity-achieving input distribution in similar scenarios is known to be discrete, with a finite number of symbols, one of which is located at the origin~\cite{abou-faycal_capacity_2001}.
            Therefore, $\mathcal{X}$ is constructed from a unipolar \acrfull{pam} with $x_1=0$, similarly to other works on the topic~\cite{jing_design_2016,marti_constellation_2024,manolakos_energy-based_2016}.

            On the other hand, the optimal decision rule~\cref{eq:ml_mu} has a major drawback which is its complexity.
            Since it increases exponentially with $K$, its practical application is restricted to systems with a small number of \acrshortpl{ue}~\cite[Sec.~16.3–2]{proakis_digital_2008}.
            A suboptimal receiver is the single-user detector that ignores the presence of other users. For the $k$-th \acrshort{ue}, it takes the form:
            \begin{equation}
                \label{eq:su_detector}
                \hat{x}_{k} = \argmin_{x\in\mathcal{X}} \herm{\mathbf{y}}\cov{\bsf{y}|x, k}^{-1}\mathbf{y} + \log|\cov{\bsf{y}|x, k}|,
            \end{equation}
            where
            \begin{equation}
                \label{eq:su_correlation}
                \cov{\bsf{y}|x, k} = |x|^2p_k\cov{\bsf{h}_k} + \cov{\bsf{z}}.
            \end{equation}
            Observe that, as expected, the single-user detector is also unable to exploit phase information.

            In general, the performance gap between \eqref{eq:ml_mu} and \eqref{eq:su_detector} is large.
            This is not, however, the case for systems with a large number of antennas (compared to the number of users) where the single-user detector achieves near-optimal performance with linear complexity, as we shall demonstrate in the sequel.

            In most works on noncoherent communications (see~\cite{cuevas_advanced_2024} and references therein) uncorrelated Rayleigh fading is assumed, \ie $\cov{\bsf{h}_k} = \id{N}$.
            However, this model can only be observed in practice if a half-wavelength-spaced uniform linear array in rich scattering is used~\cite{pizzo_fourier_2022}.
            On the other hand, the model~\cref{eq:correlation} used in this letter accounts for spatial correlation and is valid not only in the far field but also in the near field.
            We assume to have perfect knowledge of $\cov{\bsf{h}_k}$.
            In line with the \acrshort{mimo} literature at high frequencies, this knowledge can be acquired by means of a model-based acquisition method, which instead of estimating $\cov{\bsf{h}_k}$ directly, estimates its parameters and constructs the correlation matrix from them.
            In doing so, it plays a key role the model employed for its computation.
            In particular, using the far-field model would yield the following mismatched correlation matrix, even if the parameters are perfectly estimated~\cite{bacci_spherical_2023}:
            \begin{equation}
                \label{eq:su_correlation_ff}
                \cov{\bsf{y}|x, k}^{\mathrm{FF}} = |x|^2p_k\cov{\bsf{h}_k}^{\mathrm{FF}} + \cov{\bsf{z}}.
            \end{equation}
            The aim of this letter is to quantify the impact of such inaccurate modeling and to provide evidence that a noncoherent receiver employing~\cref{eq:su_correlation_ff} instead of~\cref{eq:su_correlation}, even beyond the Fraunhofer distance, significantly underestimates the achievable \acrshort{ser}.

            \subsection{Single-User Asymptotic Regime}
                \label{sec:asymptotic_regime}
                Independently of the receiver considered, a necessary condition for an arbitrarily low $\perr$ in a noncoherent system~\cite{vila-insa_singular_2025} is to have a \textit{uniquely identifiable constellation}, which in the single-user scenario\footnote{Note that if the condition is necessary for a single user it must also be necessary in the multiuser scenario.} is defined by:
    			\begin{equation*}
    				|x_a|^2\neq|x_b|^2 \iff x_a\neq x_b,\quad\forall x_a,x_b\in\mathcal{X}.
    			\end{equation*}
    			Under this hypothesis, the error probability exhibits the following properties for large $N$ and high \acrshort{snr}, respectively:
    			\begin{align}
    				\lim_{N\to\infty}\perr = 0 &\iff \lim_{N\to\infty}\rk{\cov{\bsf{h}}} = \infty,\label{eq:as_n}\\
    				\lim_{p\to\infty}\perr = 0 &\iff M=2.\label{eq:as_snr}
    			\end{align}
                Although employing large arrays enables lower-complexity detection, the channel rank is still determined by the number of scatterers: $\rk{\cov{\bsf{h}}} = L$, and it does not increase with $N$.
                Therefore,~\cref{eq:as_n} is not fulfilled and the system must exhibit an error floor.
                Regarding~\cref{eq:as_snr}, the interpretation is that, for a \acrshort{pam} of order $M=2$, the zero symbol and the non-zero symbol span different subspaces and can be detected with arbitrarily low error probability.
    			On the other hand, for $M>2$ all the non-zero symbols share the same subspace and the only vanishing pairwise error probabilities are those between $x_1=0$ and any other symbol~\cite{VilaInsa2024}.

	\section{Channel Statistics Discussion}
        \label{sec:channel_statistics}
        The performance of energy-based noncoherent systems strongly depends on the channel correlation matrix.
        For this reason, in this section we analyze the main properties of the near-field and far-field models~\cref{eq:correlation,eq:correlation_ff}.
        In doing so, we assume that scattering is local to each \acrshort{ue}, with the $k$-th \acrshort{ue} located at $\mathbf{r}_k = \trans{(r_k, \theta_k, \varphi_k)}$, and is uniformly sampled from a sphere centered at the user. We assume that the radius of the sphere is $\rho_{\mathrm{s}} = \qty{3}{\m}$ and the number of scatterers is $L=10$, unless stated otherwise.

        Antennas at the \acrshort{bs} are arranged as a \acrshort{upa} with, except where noted, $N=16\times16$ and $\Delta=\lambda$.
		The system operates at $f=\qty{30}{\GHz}$ such that the Fresnel and Fraunhofer distances are $\dist{r}=\qty{0.67}{\m}$ and $\dist{F}=\qty{10}{\m}$.

		\subsection{Spatial Correlation Spectrum}
			\label{sec:nf_ff_eigenvalues}
\begin{figure}
				\centering
				\resizebox{\columnwidth}{!}{%
					\input{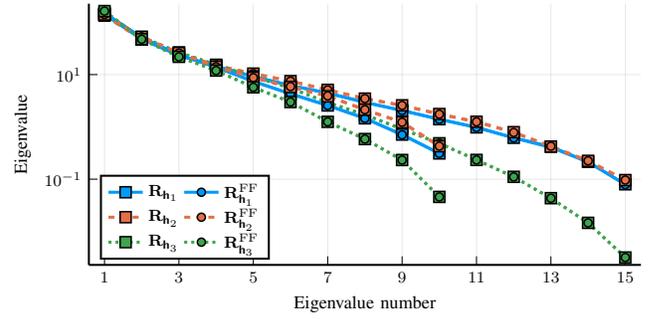}
				}
				\caption{Non-zero eigenvalues of each \acrshort{ue}'s near-field and far-field channel correlation matrices for $L=10$ and $L=15$.}
				\label{fig:eigvalues_nf}
			\end{figure}

            We consider two \acrshortpl{ue} in the near field: $\mathbf{r}_1 = (5, \qty{-30}{\degree}, \qty{-10}{\degree})$ and $\mathbf{r}_2 = (5, \qty{-20}{\degree}, \qty{0}{\degree})$, and a \acrshort{ue} in the far field, $\mathbf{r}_3 = (25, \qty{-10}{\degree}, \qty{10}{\degree})$.
			\cref{fig:eigvalues_nf} reports the eigenvalues of $\cov{\bsf{h}_1}$, $\cov{\bsf{h}_2}$ and $\cov{\bsf{h}_3}$, as obtained with \cref{eq:correlation}, and those of their respective far-field matrices, obtained through~\cref{eq:correlation_ff}.
            For all \acrshortpl{ue}, the eigenvalues of the near-field correlation matrix (rectangular markers) coincide with those of the far-field matrix (circular markers). This shows that both models exhibit the same correlation spectrum.
            Similar results have been previously observed in~\cite{demir_spatial_2024} and~\cite{dong_near-field_2022}.

            Regarding the eigenvalues of different users, we should recall that~\cref{eq:correlation,eq:correlation_ff} only depend on the position of the scatterers, but not the user.
            Thus, if two users are in the same scattering cluster the eigenvalues of their channel correlation matrices coincide, even if scatterers realizations differ.
            Indeed, for $\rho_{\mathrm{s}}=\qty{3}{\m}$ and two users located at $\mathbf{r}_1$ and $\mathbf{r}_2$, their scatterers overlap and the correlation matrices spectra are very close (see the dashed orange and solid blue lines in~\cref{fig:eigvalues_nf}).

		\subsection{Distance Between Column Spaces}
            The results in~\cref{fig:eigvalues_nf} evidence that the near-field and far-field correlation matrices from~\cref{eq:correlation,eq:correlation_ff} exhibit identical spectra, even at short communication distances.
            Therefore, the primary difference between the two models must lie in their eigenvectors, \ie $\cov{\bsf{h}_k}$ and $\cov{\bsf{h}_k}^{\mathrm{FF}}$ span different subspaces.
            However, as discussed in~\cref{sec:asymptotic_regime}, a noncoherent system such as the one here considered spans only two subspaces: the zero-symbol subspace and the non-zero-symbol subspace.
            Thus, measuring the distance between $\col(\cov{\bsf{h}_k})$ and $\col(\cov{\bsf{h}_k}^{\mathrm{FF}})$ --which asymptotically must be zero-- provides valuable insights into communication performance.
            Here, $\col(\cdot)$ denotes the subspace spanned by the matrix columns.

            There are various methods to measure the distance between subspaces, but the most commonly used in the design of Grassmannian constellations is the \textit{chordal distance},\footnote{This is also referred to as \textit{projection F-distance}, \eg \cite{edelman_geometry_1998}.} as it has a direct impact on the error probability~\cite{hochwald_systematic_2000, cuevas_advanced_2024}.
            The chordal distance is defined by:
			\begin{equation}
				\dist{ch}(\mathbf{X}, \mathbf{Y}) = \Big(\sum_{i=1}^{L}\sin^2\alpha_i\Big)^{\frac{1}{2}} = \tr{\id{}-\herm{\mathbf{Y}}\mathbf{X}\herm{\mathbf{X}}\mathbf{Y}}^{\frac{1}{2}},
			\end{equation}
			where $\mathbf{X}$ and $\mathbf{Y}$ are two $L$-rank matrices and $\alpha_i$ are the principal angles~\cite{miao_principal_1992, galantai_jordans_2006} between the subspaces spanned by $\mathbf{X}$ and $\mathbf{Y}$.
            Observe that $\dist{ch}(\mathbf{X}, \mathbf{Y}) = 0$ if and only if $\col(\mathbf{X}) = \col(\mathbf{Y})$ and $\dist{ch}(\mathbf{X}, \mathbf{Y}) = \sqrt{L}$ if and only if $ \col(\mathbf{X}) \perp \col(\mathbf{Y})$.

\begin{figure}
				\centering
				\resizebox{\columnwidth}{!}{%
					% Recommended preamble:
% \usetikzlibrary{arrows.meta}
% \usetikzlibrary{backgrounds}
% \usepgfplotslibrary{patchplots}
% \usepgfplotslibrary{fillbetween}
% \pgfplotsset{%
%     layers/standard/.define layer set={%
%         background,axis background,axis grid,axis ticks,axis lines,axis tick labels,pre main,main,axis descriptions,axis foreground%
%     }{
%         grid style={/pgfplots/on layer=axis grid},%
%         tick style={/pgfplots/on layer=axis ticks},%
%         axis line style={/pgfplots/on layer=axis lines},%
%         label style={/pgfplots/on layer=axis descriptions},%
%         legend style={/pgfplots/on layer=axis descriptions},%
%         title style={/pgfplots/on layer=axis descriptions},%
%         colorbar style={/pgfplots/on layer=axis descriptions},%
%         ticklabel style={/pgfplots/on layer=axis tick labels},%
%         axis background@ style={/pgfplots/on layer=axis background},%
%         3d box foreground style={/pgfplots/on layer=axis foreground},%
%     },
% }

\begin{tikzpicture}[/tikz/background rectangle/.style={fill={rgb,1:red,1.0;green,1.0;blue,1.0}, fill opacity={1.0}, draw opacity={1.0}}, show background rectangle]
\begin{axis}[point meta max={nan}, point meta min={nan}, legend cell align={left}, legend columns={1}, title={}, title style={at={{(0.5,1)}}, anchor={south}, font={{\fontsize{14 pt}{18.2 pt}\selectfont}}, color={rgb,1:red,0.0;green,0.0;blue,0.0}, draw opacity={1.0}, rotate={0.0}, align={center}}, legend style={color={rgb,1:red,0.0;green,0.0;blue,0.0}, draw opacity={1.0}, line width={1}, solid, fill={rgb,1:red,1.0;green,1.0;blue,1.0}, fill opacity={1.0}, text opacity={1.0}, font={{\fontsize{8 pt}{10.4 pt}\selectfont}}, text={rgb,1:red,0.0;green,0.0;blue,0.0}, cells={anchor={center}}, at={(0.02, 0.02)}, anchor={south west}}, axis background/.style={fill={rgb,1:red,1.0;green,1.0;blue,1.0}, opacity={1.0}}, anchor={north west}, xshift={1.0mm}, yshift={-1.0mm}, width={112.3mm}, height={61.5mm}, scaled x ticks={false}, xlabel={Distance [m]}, x tick style={color={rgb,1:red,0.0;green,0.0;blue,0.0}, opacity={1.0}}, x tick label style={color={rgb,1:red,0.0;green,0.0;blue,0.0}, opacity={1.0}, rotate={0}}, xlabel style={at={(ticklabel cs:0.5)}, anchor=near ticklabel, at={{(ticklabel cs:0.5)}}, anchor={near ticklabel}, font={{\fontsize{9 pt}{11.700000000000001 pt}\selectfont}}, color={rgb,1:red,0.0;green,0.0;blue,0.0}, draw opacity={1.0}, rotate={0.0}}, xmode={log}, log basis x={10}, xmajorgrids={true}, xmin={0.8530394184621627}, xmax={234.45575394458902}, xticklabels={{$10^0$,$10^{2}$,$10^{3}$,$d_{\mathrm{F}_{1}}$,$d_{\mathrm{F}_{2}}$,$d_{\mathrm{F}_{3}}$}}, xtick={{1.0,100.0,1000.0,10.239999771118164,23.040000915527344,40.959999084472656}}, xtick align={inside}, xticklabel style={font={{\fontsize{8 pt}{10.4 pt}\selectfont}}, color={rgb,1:red,0.0;green,0.0;blue,0.0}, draw opacity={1.0}, rotate={0.0}}, x grid style={color={rgb,1:red,0.0;green,0.0;blue,0.0}, draw opacity={0.1}, line width={0.5}, solid}, axis x line*={left}, x axis line style={color={rgb,1:red,0.0;green,0.0;blue,0.0}, draw opacity={1.0}, line width={1}, solid}, scaled y ticks={false}, ylabel={$d_{\mathrm{ch}}(\mathbf{U}_{\bsf{h}}, \mathbf{U}_{\bsf{h}}^{\mathrm{FF}})/\sqrt{L}$}, y tick style={color={rgb,1:red,0.0;green,0.0;blue,0.0}, opacity={1.0}}, y tick label style={color={rgb,1:red,0.0;green,0.0;blue,0.0}, opacity={1.0}, rotate={0}}, ylabel style={at={(ticklabel cs:0.5)}, anchor=near ticklabel, at={{(ticklabel cs:0.5)}}, anchor={near ticklabel}, font={{\fontsize{9 pt}{11.700000000000001 pt}\selectfont}}, color={rgb,1:red,0.0;green,0.0;blue,0.0}, draw opacity={1.0}, rotate={0.0}}, ymode={log}, log basis y={10}, ymajorgrids={true}, ymin={0.009286884348304101}, ymax={1.1373792077826541}, yticklabels={{$10^{-2}$,$10^{-1}$,$10^{0}$}}, ytick={{0.01,0.1,1.0}}, ytick align={inside}, yticklabel style={font={{\fontsize{8 pt}{10.4 pt}\selectfont}}, color={rgb,1:red,0.0;green,0.0;blue,0.0}, draw opacity={1.0}, rotate={0.0}}, y grid style={color={rgb,1:red,0.0;green,0.0;blue,0.0}, draw opacity={0.1}, line width={0.5}, solid}, axis y line*={left}, y axis line style={color={rgb,1:red,0.0;green,0.0;blue,0.0}, draw opacity={1.0}, line width={1}, solid}, colorbar={false}]
    \addplot[color={rgb,1:red,0.0;green,0.6056;blue,0.9787}, name path={5fe6d407-2bdc-4efe-8874-53aa5c21f2c5}, draw opacity={1.0}, line width={1.5}, solid, mark={*}, mark size={2.625 pt}, mark repeat={1}, mark options={color={rgb,1:red,0.0;green,0.0;blue,0.0}, draw opacity={1.0}, fill={rgb,1:red,0.0;green,0.6056;blue,0.9787}, fill opacity={1.0}, line width={0.75}, rotate={0}, solid}]
        table[row sep={\\}]
        {
            \\
            1.0  0.878987650450889  \\
            1.801648230654411  0.8394474194472179  \\
            3.24593634702017  0.7027029608480638  \\
            5.848035476425733  0.4933774498996418  \\
            10.536102768906645  0.2751163670185473  \\
            18.982350911593706  0.13557380391119495  \\
            34.19951893353395  0.060900645862369394  \\
            61.61550277583347  0.03503949676114714  \\
            111.00946155696226  0.018422488980417477  \\
            200.00000000000003  0.010640581792137595  \\
        }
        ;
    \addlegendentry {$N=16 \times 16$}
    \addplot[color={rgb,1:red,0.0;green,0.6056;blue,0.9787}, name path={b19e3a40-3b3b-488e-80e6-82733e3e6d6e}, draw opacity={1.0}, line width={1.5}, solid, forget plot]
        table[row sep={\\}]
        {
            \\
            10.240000000000004  7.582890588171957e-5  \\
            10.240000000000004  139.29660516689108  \\
        }
        ;
    \addplot[color={rgb,1:red,0.8889;green,0.4356;blue,0.2781}, name path={8fdf6109-b7f4-4f79-8df7-c6b35a856c5c}, draw opacity={1.0}, line width={1.5}, dashed, mark={square*}, mark size={2.625 pt}, mark repeat={1}, mark options={color={rgb,1:red,0.0;green,0.0;blue,0.0}, draw opacity={1.0}, fill={rgb,1:red,0.8889;green,0.4356;blue,0.2781}, fill opacity={1.0}, line width={0.75}, rotate={0}, solid}]
        table[row sep={\\}]
        {
            \\
            1.0  0.9860620797345752  \\
            1.801648230654411  0.985912339139477  \\
            3.24593634702017  0.9658979089832072  \\
            5.848035476425733  0.845147082385007  \\
            10.536102768906645  0.6045759063504075  \\
            18.982350911593706  0.3279074766230032  \\
            34.19951893353395  0.16291249074028186  \\
            61.61550277583347  0.07449367046540985  \\
            111.00946155696226  0.040052454706434246  \\
            200.00000000000003  0.021714379775093585  \\
        }
        ;
    \addlegendentry {$N=24 \times 24$}
    \addplot[color={rgb,1:red,0.8889;green,0.4356;blue,0.2781}, name path={36c80f89-0641-4a6d-ae8d-dbc3969763b2}, draw opacity={1.0}, line width={1.5}, dashed, forget plot]
        table[row sep={\\}]
        {
            \\
            23.039999999999996  7.582890588171957e-5  \\
            23.039999999999996  139.29660516689108  \\
        }
        ;
    \addplot[color={rgb,1:red,0.2422;green,0.6433;blue,0.3044}, name path={4982325a-7e6f-4ba8-9c68-78fe09b43137}, draw opacity={1.0}, line width={1.5}, dotted, mark={diamond*}, mark size={2.625 pt}, mark repeat={1}, mark options={color={rgb,1:red,0.0;green,0.0;blue,0.0}, draw opacity={1.0}, fill={rgb,1:red,0.2422;green,0.6433;blue,0.3044}, fill opacity={1.0}, line width={0.75}, rotate={0}, solid}]
        table[row sep={\\}]
        {
            \\
            1.0  0.9926815440343797  \\
            1.801648230654411  0.9923614360981835  \\
            3.24593634702017  0.9913601738604897  \\
            5.848035476425733  0.9777517667050155  \\
            10.536102768906645  0.879887612587241  \\
            18.982350911593706  0.5860783299376617  \\
            34.19951893353395  0.3080931104685158  \\
            61.61550277583347  0.13901821161961392  \\
            111.00946155696226  0.07031313188702006  \\
            200.00000000000003  0.038905047027445204  \\
        }
        ;
    \addlegendentry {$N=32 \times 32$}
    \addplot[color={rgb,1:red,0.2422;green,0.6433;blue,0.3044}, name path={dea0aa46-649f-4d43-bf43-b13a176e5480}, draw opacity={1.0}, line width={1.5}, dotted, forget plot]
        table[row sep={\\}]
        {
            \\
            40.96000000000002  7.582890588171957e-5  \\
            40.96000000000002  139.29660516689108  \\
        }
        ;
\end{axis}
\end{tikzpicture}
				}
				\caption{Chordal distance between $\col(\cov{\bsf{h}})$ and $\col(\cov{\bsf{h}}^{\mathrm{FF}})$ for $N=16\times16$, $N=24\times24$ and $N=32\times32$, in terms of the physical distance between transmitter and receiver. The Fraunhofer distances for each case are indicated as $\dist{F_1}$, $\dist{F_2}$ and $\dist{F_3}$.}
				\label{fig:subspaces_distance}
			\end{figure}
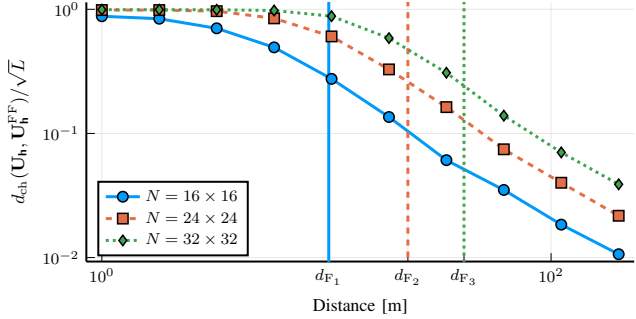

            Now consider a single \acrshort{ue} moving in a straight line from $\trans{(1, \qty{-30}{\degree}, \qty{-10}{\degree})}$ to $\trans{(200, \qty{-0.15}{\degree}, \qty{-10}{\degree})}$ under three array configurations: $N=16\times16$, $N=24\times24$ and $N=32\times32$.
            In~\cref{fig:subspaces_distance}, we represent $\dist{ch}(\mathbf{U}_{\bsf{h}},\mathbf{U}_{\bsf{h}}^{\mathrm{FF}})/\sqrt{L}$, where $\mathbf{U}_{\bsf{h}},\mathbf{U}_{\bsf{h}}^{\mathrm{FF}}\in\cnumbers^{N\times L}$ are the matrices containing the $L$ eigenvectors associated to the non-zero eigenvalues of $\cov{\bsf{h}}$ and $\cov{\bsf{h}}^{\mathrm{FF}}$ (the \acrshort{ue} subindex has been omitted for simplicity).
            Differently from the eigenvalues, which were already equal at $r=\dist{r}$, the normalized chordal distance starts being one, that is, the subspaces spanned by $\cov{\bsf{h}}$ and $\cov{\bsf{h}}^{\mathrm{FF}}$ are orthogonal.
            Although $\dist{ch}(\mathbf{U}_{\bsf{h}},\mathbf{U}_{\bsf{h}}^{\mathrm{FF}})/\sqrt{L}$ decreases with the distance between transmitter and receiver, at the Fraunhofer distance it still is a $30 \%$ of its maximum value, and keeps decreasing at the same rate, suggesting that it is far from converging.
            In practice, this means that employing the far-field model makes the receiver operate in an appreciably mismatched mode, degrading the performance as it will be shown later.

            On the other hand, incrementing the number of antennas (and hence the array aperture) also increases the chordal distance.
            Indeed, for $N\to\infty \implies \dist{ch}(\mathbf{U}_{\bsf{h}}, \mathbf{U}_{\bsf{h}}^{\mathrm{FF}}) = \sqrt{L}$, and thus $\col(\cov{\bsf{h}}) \perp \col(\cov{\bsf{h}}^{\mathrm{FF}})$.
            This is a specific instance of a well-known result in the large-array literature, namely the asymptotic orthogonality of near-field beamfocusing vectors~\cite{wu_multiple_2023, ramezani_near-field_2024}. To the best of the authors’ knowledge, this orthogonality has only been exploited in coherent communication systems. However, it can also be utilized in noncoherent systems to multiplex different \acrshortpl{ue}.

	\section{Numerical Results}
        In order to verify the impact down to detection of a receiver employing the mismatched model~\cref{eq:su_correlation_ff} as well as the performance of the suboptimal single-user detector~\cref{eq:su_detector}, we evaluate the error probability of a system employing the 4-\acrshort{pam} constellation that minimizes the \acrshort{ser} of the optimal quadratic detector (see~\cite{marti_constellation_2024,VilaInsa2024} for further details).\footnote{To focus on receiver design, the constellation has been optimized for each~$N$ and each realization of scatterers location. The design is the same for near-field and far-field models since they share the same eigenvalues.}
        Specifically, to assess the mismatched mode, we consider a single \acrshort{ue} moving away from the \acrshort{bs}.
        On the other hand, the performance of the suboptimal detector is analyzed in a scenario involving five static \acrshortpl{ue} simultaneously communicating with the same \acrshort{bs}.

        \subsection{Single-User Mismatched Mode}
            \label{sec:mismatched_mode}
            Let's start with a single \acrshort{ue} moving along a straight line from $\trans{(1, \qty{-30}{\degree}, \qty{-10}{\degree})}$ to $\trans{(200, \qty{-0.15}{\degree}, \qty{-10}{\degree})}$.
            The \acrshort{bs} implements the receiver from~\cref{eq:su_detector}, that for $K=1$ is the \acrshort{ml} detector, to decide the transmitted symbol.

\begin{figure}
				\centering
				\resizebox{\columnwidth}{!}{%
					\input{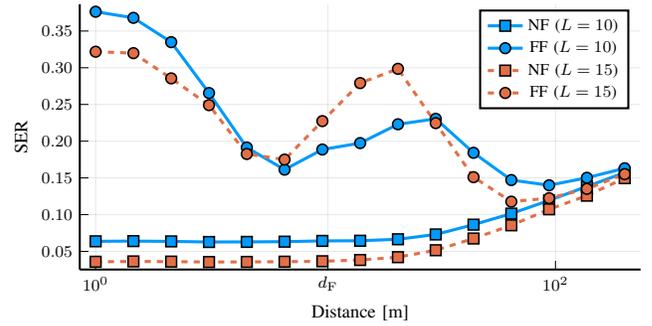}
				}
				\caption{Error probability of a \acrshort{ml} detector employing the exact or mismatched model at different distances. The simulation is performed for $\mathrm{SNR}=\qty{20}{\decibel}$.}
				\label{fig:ser_ff_nf_distance}
			\end{figure}

            To show the impact of operating in a mismatched mode, in~\cref{fig:ser_ff_nf_distance} we illustrate the \acrshort{ser} of the proposed setup both for a receiver employing the exact correlation model and the mismatched one, for $L=10$ and $L=15$.
            Besides the severe difference between \acrshortpl{ser} until $10\dist{F}$, using the far-field model may result in unexpected performance behaviors due to the mismatch.
            Regarding the increase of error probability with distance, even when employing the exact model, it is caused by the loss of resolution between scatterers in the local scattering model.
            That is, when the distance increases, all signals paths tend to come from the same angle and the rank of the channel correlation matrix goes to one.

            The conclusions of this analysis are twofold: first, a receiver should always utilize the exact model, even beyond the Fraunhofer distance; second, noncoherent communication systems can take advantage from near field propagation conditions.

        \subsection{Multiuser Detection}
            Consider a scenario with five \acrshortpl{ue} arbitrarily located at positions: $\mathbf{r}_1 = \trans{(5, \qty{-30}{\degree}, \qty{-20}{\degree})}$, $\mathbf{r}_2 = \trans{(10, \qty{-25}{\degree}, \qty{-10}{\degree})}$, $\mathbf{r}_3 = \trans{(15, \qty{-20}{\degree}, \qty{0}{\degree})}$, $\mathbf{r}_4 = \trans{(20, \qty{-15}{\degree}, \qty{10}{\degree})}$ and $\mathbf{r}_5 = \trans{(25, \qty{-10}{\degree}, \qty{20}{\degree})}$.
            In order to multiplex them, scattering clusters must be separated, so throughout this section we consider $\rho_{\mathrm{s}}=\qty{1}{\m}$ to avoid overlapping.
            The receiver implements the detector from~\cref{eq:su_detector}, and its performance with different number of antennas is compared with the scenario where only the \acrshort{ue} of interest is present.
            For the sake of completeness, we also consider the mismatched mode case.

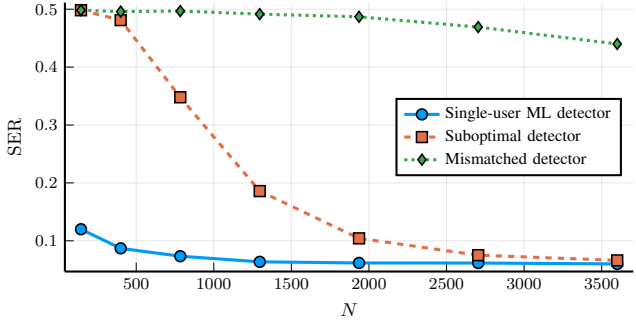
\begin{figure}
                \centering
                \resizebox{\columnwidth}{!}{%
					% Recommended preamble:
% \usetikzlibrary{arrows.meta}
% \usetikzlibrary{backgrounds}
% \usepgfplotslibrary{patchplots}
% \usepgfplotslibrary{fillbetween}
% \pgfplotsset{%
%     layers/standard/.define layer set={%
%         background,axis background,axis grid,axis ticks,axis lines,axis tick labels,pre main,main,axis descriptions,axis foreground%
%     }{
%         grid style={/pgfplots/on layer=axis grid},%
%         tick style={/pgfplots/on layer=axis ticks},%
%         axis line style={/pgfplots/on layer=axis lines},%
%         label style={/pgfplots/on layer=axis descriptions},%
%         legend style={/pgfplots/on layer=axis descriptions},%
%         title style={/pgfplots/on layer=axis descriptions},%
%         colorbar style={/pgfplots/on layer=axis descriptions},%
%         ticklabel style={/pgfplots/on layer=axis tick labels},%
%         axis background@ style={/pgfplots/on layer=axis background},%
%         3d box foreground style={/pgfplots/on layer=axis foreground},%
%     },
% }

\begin{tikzpicture}[/tikz/background rectangle/.style={fill={rgb,1:red,1.0;green,1.0;blue,1.0}, fill opacity={1.0}, draw opacity={1.0}}, show background rectangle]
\begin{axis}[point meta max={nan}, point meta min={nan}, legend cell align={left}, legend columns={1}, title={}, title style={at={{(0.5,1)}}, anchor={south}, font={{\fontsize{14 pt}{18.2 pt}\selectfont}}, color={rgb,1:red,0.0;green,0.0;blue,0.0}, draw opacity={1.0}, rotate={0.0}, align={center}}, legend style={color={rgb,1:red,0.0;green,0.0;blue,0.0}, draw opacity={1.0}, line width={1}, solid, fill={rgb,1:red,1.0;green,1.0;blue,1.0}, fill opacity={1.0}, text opacity={1.0}, font={{\fontsize{8 pt}{10.4 pt}\selectfont}}, text={rgb,1:red,0.0;green,0.0;blue,0.0}, cells={anchor={center}}, at={(0.98, 0.5)}, anchor={east}}, axis background/.style={fill={rgb,1:red,1.0;green,1.0;blue,1.0}, opacity={1.0}}, anchor={north west}, xshift={1.0mm}, yshift={-1.0mm}, width={112.3mm}, height={61.5mm}, scaled x ticks={false}, xlabel={$N$}, x tick style={color={rgb,1:red,0.0;green,0.0;blue,0.0}, opacity={1.0}}, x tick label style={color={rgb,1:red,0.0;green,0.0;blue,0.0}, opacity={1.0}, rotate={0}}, xlabel style={at={(ticklabel cs:0.5)}, anchor=near ticklabel, at={{(ticklabel cs:0.5)}}, anchor={near ticklabel}, font={{\fontsize{9 pt}{11.700000000000001 pt}\selectfont}}, color={rgb,1:red,0.0;green,0.0;blue,0.0}, draw opacity={1.0}, rotate={0.0}}, xmajorgrids={true}, xmin={40.319999999999936}, xmax={3703.6800000000003}, xticklabels={{$500$,$1000$,$1500$,$2000$,$2500$,$3000$,$3500$}}, xtick={{500.0,1000.0,1500.0,2000.0,2500.0,3000.0,3500.0}}, xtick align={inside}, xticklabel style={font={{\fontsize{8 pt}{10.4 pt}\selectfont}}, color={rgb,1:red,0.0;green,0.0;blue,0.0}, draw opacity={1.0}, rotate={0.0}}, x grid style={color={rgb,1:red,0.0;green,0.0;blue,0.0}, draw opacity={0.1}, line width={0.5}, solid}, axis x line*={left}, x axis line style={color={rgb,1:red,0.0;green,0.0;blue,0.0}, draw opacity={1.0}, line width={1}, solid}, scaled y ticks={false}, ylabel={$\mathrm{SER}$}, y tick style={color={rgb,1:red,0.0;green,0.0;blue,0.0}, opacity={1.0}}, y tick label style={color={rgb,1:red,0.0;green,0.0;blue,0.0}, opacity={1.0}, rotate={0}}, ylabel style={at={(ticklabel cs:0.5)}, anchor=near ticklabel, at={{(ticklabel cs:0.5)}}, anchor={near ticklabel}, font={{\fontsize{9 pt}{11.700000000000001 pt}\selectfont}}, color={rgb,1:red,0.0;green,0.0;blue,0.0}, draw opacity={1.0}, rotate={0.0}}, ymode={log}, log basis y={10}, ymajorgrids={true}, ymin={0.0001626300935649449}, ymax={0.6281432064045231}, yticklabels={{$10^{-3}$,$10^{-2}$,$10^{-1}$}}, ytick={{0.001,0.01,0.1}}, ytick align={inside}, yticklabel style={font={{\fontsize{8 pt}{10.4 pt}\selectfont}}, color={rgb,1:red,0.0;green,0.0;blue,0.0}, draw opacity={1.0}, rotate={0.0}}, y grid style={color={rgb,1:red,0.0;green,0.0;blue,0.0}, draw opacity={0.1}, line width={0.5}, solid}, axis y line*={left}, y axis line style={color={rgb,1:red,0.0;green,0.0;blue,0.0}, draw opacity={1.0}, line width={1}, solid}, colorbar={false}]
    \addplot[color={rgb,1:red,0.0;green,0.6056;blue,0.9787}, name path={4}, draw opacity={1.0}, line width={1.5}, solid, mark={*}, mark size={2.625 pt}, mark repeat={1}, mark options={color={rgb,1:red,0.0;green,0.0;blue,0.0}, draw opacity={1.0}, fill={rgb,1:red,0.0;green,0.6056;blue,0.9787}, fill opacity={1.0}, line width={0.75}, rotate={0}, solid}]
        table[row sep={\\}]
        {
            \\
            144.0  0.017692727272727273  \\
            400.0  0.0037345454545454544  \\
            784.0  0.0014181818181818182  \\
            1024.0  0.001114  \\
            1296.0  0.0008672727272727273  \\
            1936.0  0.0005345454545454545  \\
            2704.0  0.00034  \\
            3600.0  0.00020545454545454545  \\
        }
        ;
    \addlegendentry {Single-user ML detector}
    \addplot[color={rgb,1:red,0.8889;green,0.4356;blue,0.2781}, name path={5}, draw opacity={1.0}, line width={1.5}, dashed, mark={square*}, mark size={2.625 pt}, mark repeat={1}, mark options={color={rgb,1:red,0.0;green,0.0;blue,0.0}, draw opacity={1.0}, fill={rgb,1:red,0.8889;green,0.4356;blue,0.2781}, fill opacity={1.0}, line width={0.75}, rotate={0}, solid}]
        table[row sep={\\}]
        {
            \\
            144.0  0.49721454545454546  \\
            400.0  0.30338363636363636  \\
            784.0  0.08290545454545455  \\
            1024.0  0.003462  \\
            1296.0  0.0012654545454545455  \\
            1936.0  0.0005181818181818181  \\
            2704.0  0.00034545454545454544  \\
            3600.0  0.00020727272727272727  \\
        }
        ;
    \addlegendentry {Suboptimal detector}
    \addplot[color={rgb,1:red,0.2422;green,0.6433;blue,0.3044}, name path={6}, draw opacity={1.0}, line width={1.5}, dotted, mark={diamond*}, mark size={2.625 pt}, mark repeat={1}, mark options={color={rgb,1:red,0.0;green,0.0;blue,0.0}, draw opacity={1.0}, fill={rgb,1:red,0.2422;green,0.6433;blue,0.3044}, fill opacity={1.0}, line width={0.75}, rotate={0}, solid}]
        table[row sep={\\}]
        {
            \\
            144.0  0.4970127272727273  \\
            400.0  0.40385272727272725  \\
            784.0  0.4624309090909091  \\
            1024.0  0.458088  \\
            1296.0  0.44028  \\
            1936.0  0.3516090909090909  \\
            2704.0  0.17906  \\
            3600.0  0.07218909090909091  \\
        }
        ;
    \addlegendentry {Mismatched detector}
\end{axis}
\end{tikzpicture}
				}
                \caption{Error probability of the single-user detector in a multiuser environment for increasing number of antennas. The single-user \acrshort{ml} and the mismatched receivers are depicted for comparison and completeness. The \acrshort{sinr} is $\qty{20}{\decibel}$.}
                \label{fig:ser_multiuser}
            \end{figure}

            The results of the aforementioned experiment for the fourth \acrshort{ue} are shown in~\cref{fig:ser_multiuser} (other \acrshortpl{ue} are omitted due to the symmetry of the problem).
            As expected, the \acrshort{ser} of the single-user detector tends to that of the optimal detector when the number of antennas increases.
            In particular, for $N\geq1500$ both receivers exhibit almost the same performance.
            On the other hand, the \acrshort{ser} of the mismatched receiver is always high, independently of the number of antennas.

	\section{Conclusions}
        In this paper, we analyzed the uplink of a \acrshort{nlos} one-shot noncoherent communication system, with arbitrarily located single-antenna \acrshortpl{ue}, and a \acrshort{bs} equipped with a large antenna array operating at the mmWave band.
        Since the performance of energy-based noncoherent systems is tightly linked to the channel correlation matrices, we examined both exact (\ie near-field) and far-field models of the channel correlation matrix.
        In particular, we showed that the subspaces spanned by these models differ significantly, even when the distance between the considered \acrshort{ue} and the \acrshort{bs} is greater than the Fraunhofer distance.
        More importantly, we also illustrated that the aforementioned discrepancy clearly translates as well to a significant \acrshort{ser} degradation at the receiver side.
        Furthermore, we proved that by exploiting large arrays it is possible to multiplex different \acrshortpl{ue}, even without instantaneous \acrshort{csi}, and achieve near-optimal performance with linear complexity (in the number of users) using a single-user detector.
        This result suggests that constellations designed for a single user can also be effectively used in a multiuser setting, provided that the \acrshort{bs} has a sufficiently large number of antennas.

	\bibliography{references}
\end{document}